\documentclass{sf2a-conf2012}
\usepackage{graphicx}
\usepackage{hyperref}
\usepackage[]{natbib}  
\usepackage[cyr]{aeguill}
\usepackage{epstopdf}

\def\BibTeX{{\rm B\kern-.05em{\sc i\kern-.025em b}\kern-.08em
    T\kern-.1667em\lower.7ex\hbox{E}\kern-.125emX}}
\bibpunct{(}{)}{;}{a}{}{,}  

\begin{document}

\TitreGlobal{SF2A 2012}


\title{A possible impact near the Milky Way of a former major merger in the Local Group}

\runningtitle{Result of the M31 major merger}

\author{S. Fouquet}\address{Univ Paris Diderot, Sorbonne Paris Cit\'e, GEPI, UMR 8111, F-75205 Paris, France}

\author{F. Hammer}\address{Laboratoire GEPI, Observatoire de Paris, CNRS-UMR8111, Univ Paris Diderot, Sorbonne Paris Cit\'e, 5 place Jules Janssen, 92195 Meudon France}

\author{Y. Yang $^{1,}$}\address{National Astronomical Observatoires, Chinese Academy of Sciences, 20A Datun Road, Chaoyang District, Beijing 100012, China}

\author{M. Puech$^2$}

\author{H. Flores$^2$}


\setcounter{page}{237}


\maketitle


\begin{abstract}
The Milky Way (MW) dwarf system presents two exceptional features, namely it forms a thick plane called the Vast Polar Structure (VPOS), and the two biggest dwarves, the Magellanic Clouds (MCs), are irregular galaxies that are almost never seen at such a proximity from a luminous, L* galaxy. Investigating from our modelling of M31 as a result of a former gas-rich major merger, we find that one of the expected tidal tail produced during the event may have reached the MW. Such a coincidence may appear quite exceptional, but the MW indeed lies within the small volume delineated by the tidal tail at the present epoch. 

In our scenario, most of the MW dwarves, including the MCs, may have been formed within a tidal tail formed during the former merger in the Local Group. It leads to a fair reproduction of the VPOS as well as to a simple explanation of the MCs proximity to the MW, i.e. accounting for both exceptional features of the MW dwarf distributions. However this scenario predicts dark-matter free MW dwarves, which is in apparent contradiction with their intrinsically large velocity dispersions. To be established or discarded, this requires to further investigate their detailed interactions with the MW potential.




\end{abstract}



\section{Introduction}

The anisotropy of the MW satellites spatial distribution had led \cite{Lynden-Bell1976} to suggest that the MCs, Draco, Scuptor and Ursa Minor may form a stream. More recently, \cite{Kroupa2005} and \cite{Metz2007, Metz2009,Fouquet2012} found that the classical dwarf galaxies have their locations and orbital motions inscribed into a thick plane, named the VPOS \citep{Pawlowski2012}, which could provide an important challenge for the $\Lambda$CDM cosmological simulations \citep{Kroupa2005}. Realisations of recent simulations show that the VPOS is much thinner than expected \citep{Wang2012}, and this leads to a general consensus that progenitors of MW dwarves should have reached the MW in an organised motion, either from a very compact group or from a tidal tail \citep{Pawlowski2011}. Both issues have their pros and cons, because such a compact group seems unlikely \citep{Metz2009} or because tidal dwarves (TDGs) are dark-free in contradiction with measurements of their velocity dispersion  \citep{Walker2009}.


Perhaps the above problem is related to the MCs proximity to the MW (50kpc), which is also an enigma given the fact that no similar configuration is found in the nearby Universe. Indeed, the probability to find two massive ($>$ 10$^8$ M$_{\odot}$) dwarf irregular galaxies close (< 60 kpc) to their host L* galaxies is very small \citep[$< 0.4$ \%,][]{Robotham2012}, and the only examples found are in groups made by two L* galaxies, such as M31 and the MW in the Local Group.

 The present work investigates a new scenario linking together the formation of the VPOS and the position of the MCs. Following \cite{Pawlowski2011}, we suggest that the 11 classical dwarves could be ancient TDGs, formed due to a major merger. The difference with the \cite{Pawlowski2011}' scenario comes from our assumption that it is linked to the M31 tumultuous past history, instead of that of the MW. Our scenario investigates whether the VPOS could be the result of the interaction between a tidal tail linking M31 to the MW, while the exceptional MCs proximity to the MW would simply reflect the tiny probability for a tidal tail send by M31 to interact with the MW.

\section{TDGs within a tidal tail ejected by M31 towards the MW}

According to \cite{Hammer2010}, M31 could be the result of a gas-rich major merger (mass ratio $\sim 3 \pm 0.5:1$, r$_{pericenter}$ $\sim 25 \pm 5$), with a first passage 8-9 Gyrs ago and a fusion time 5-6 Gyr ago, providing the formation of several tidal tails. At the present-time in the simulations, the tidal tail generated during the first passage is presently long enough (>1.5 Mpc) to reach the Milky Way (785 kpc from M31). However, it could have been ejected towards different directions. In fact, the angular momentum of a tidal tail in a major merger follow the orbital angular momentum which is within $\pm$ 25$^{\circ}$ to that of the remnant disk. As the M31 disk is nearly edge-on (77$^{\circ}$) for a MW observer, the tidal tail must be in a thick plane aligned to the M31 disk and that includes the MW. An additional constrain is provided by the Giant Stream (with an angular uncertainty of 20$^{\circ}$) that is also reproduced by the \cite{Hammer2010} simulations. Consequently, the tidal tail must lie in a quite small solid angle representing only  5\% of the 4$\pi$ steradian sphere (see Fig. 1). It is quite an exceptional coincidence \citep{Fouquet2012} that the MW also lies in this small solid angle (see Fig. 1, right panel), that strengthens the possibility of an encounter between the MW and a tidal tail originated from M31. Perhaps this is a reminiscence of the investigations  of \cite{Robotham2012} in the local volume.

\begin{figure}[ht!]
 \centering
 \includegraphics[width=0.8\textwidth,clip]{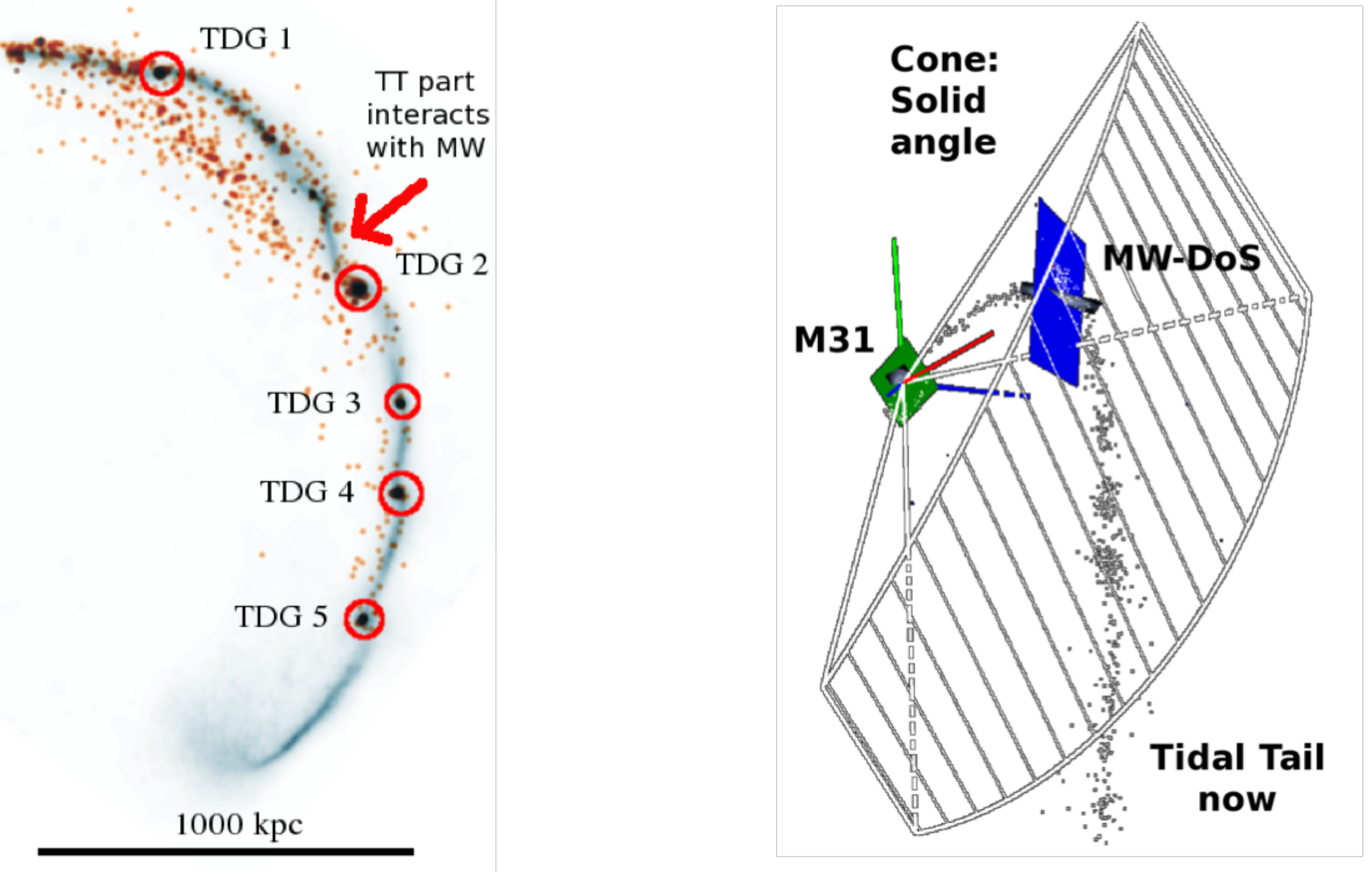}
  \caption{\textit{Left panel:} An example of first tidal tail at the present-time with five formed TDGs. The blue color code for the gas density, and the orange one for the stars. \textit{Right panel:} the tidal tail interacting with the MW. The solid angle represents the space where the tidal tail is constrained to move.}
  \label{author1:fig1}
\end{figure}

The left panel of Fig. 1 shows the tidal tail which could interact with MW at the present time. It is extracted from a GADGET2 simulation whose the number of particles exceeds 2 millions. The five overdensities are TDGs resulting of gas and baryonic matter collapse due to gravitational instability and gas cooling \citep{Wetzstein2007}. Their baryonic masses are larger than $10^8$ M$_{\odot}$ and for the most massive one (TDG 2), its mass is close to $10^9$ M$_{\odot}$, one third of the baryonic LMC mass. The high mass of the TDG2 and its position at the center of the tidal tail and not at the top of it is apparently discrepant from the \cite{Bournaud2006} results. This could be caused by the high fraction of gas of the progenitors galaxies (>60\%) that is assumed in our study, which make easier the formation of high mass TDG formation \citep{Wetzstein2007}.

\section{Reproduction of the VPOS by the MW-tidal tail encounter}

From the work made by Hammer et al. (2010), we confirm that a tidal tail including several TDGs, could interact with the MW. Then, we have developped a simple formalism to simulate the interaction between the simulated tidal tail with a MW analytic mass profile model to verify whether or not it could reproduce the VPOS \citep{Fouquet2012}.

We have first tried to match the LMC position and its trajectory to that of the tidal tail. We have treated the LMC case alone because its position and proper velocity are the best known among the MW dwarf galaxies. Tracing back the LMC depends on the current position and velocity of the LMC, the MW and M31, and on their total masses \citep{Yang2010}. Observations let significant freedom in establishing the total mass of massive galaxies and even more for the M31 transversal motion. Here we adopt $M_{baryonic}/M_{DM}$ = 20\%, instead of half this value from \cite{Yang2010}. We derive proper motions for which the LMC trajectory and velocity match well the trajectory of the tidal tail for M31, assuming a M31 motion close to that derived by \cite{VdM2012}. 

Then, we have performed N-body simulations to investigate the encountering between the tidal tail and the MW, leading to a reasonable fit of both spatial and orbital motion distributions of the dwarf galaxies that lie in the VPOS, within the measurement uncertainties (see Fig. 2 and \cite{Fouquet2012} for the orbital motions).

\begin{figure}[ht!]
 \centering
 \includegraphics[width=1\textwidth,clip]{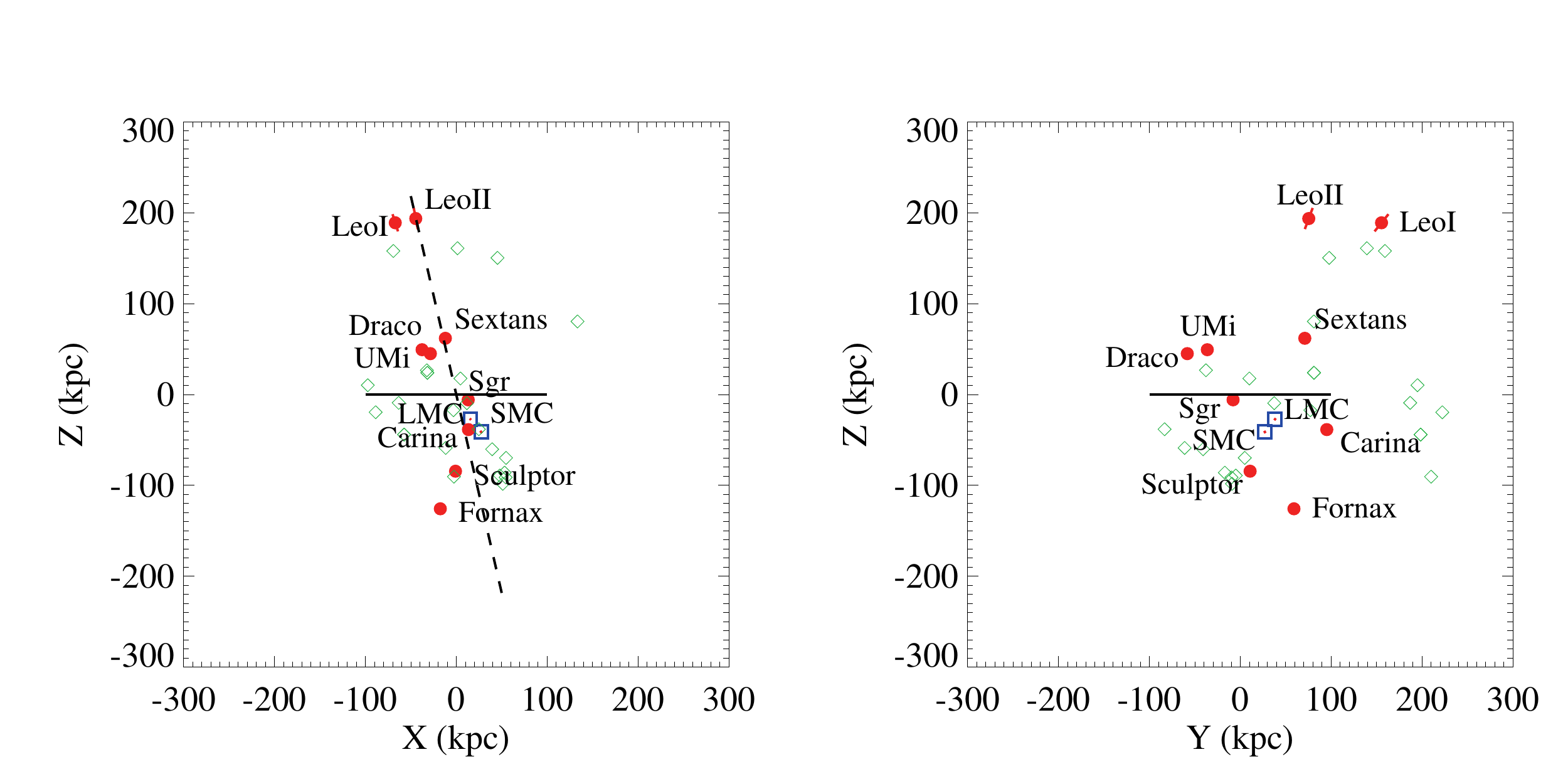}
  \caption{Spatial distribution of the dwarf galaxies and of the simulated particles (the green diamonds). The blue squares represent the Magellanic Clouds, the red dots the dwarf spheroidals and the red lines their uncertainties. On the two panels, the MW (the solide line) is viewed edge-on. \textbf{Left:} the VPOS (the dashed line) is viewed edge-on. \textbf{Right:} the VPOS is rotated from the position of the left panel by 90$^{\circ}$ along the MW disk axis.}
  \label{author1:fig1}
\end{figure}

\section{Conclusion and prospectives}

The suggested scenario has the advantages to be consistent with the hierarchical model, for which a significant part of gas-rich major mergers forming disks have occured at z $\sim$ 0.4-0.8 \cite{Hammer2009}, as it could be also the case for M31. Moreover it naturally explains within a single scheme the two exceptional features in MW dwarf galaxies, i.e., the VPOS and the proximity of the MCs to the MW.

However, it remains speculative with respect to the large velocity dispersions of the MW dwarf galaxies. Indeed, if they were formed in a tidal tail, they must contain a very small fraction of dark matter providing small M/L ratio, while their M/L ratio deduced from their velocity dispersions are reaching very large values \citep{Walker2009}. Perhaps the MW dwarf galaxies are not virialised and former numerical studies have shown that a dwarf galaxy in interaction with its host galaxy could show apparently large M/L values if it is assumed stable whereas it is not \citep{Kroupa1997}. Further studies are needed to investigate this important issue as well as to understand how the gas may have been stripped during such an interaction to form spheroidal dwarves .

\bibliographystyle{aa}  
\bibliography{sf2a} 

\begin{thebibliography}{17}
\expandafter\ifx\csname natexlab\endcsname\relax\def\natexlab#1{#1}\fi

\bibitem[{{Bournaud} \& {Duc}(2006)}]{Bournaud2006}
{Bournaud}, F. \& {Duc}, P.-A. 2006, \aap, 456, 481

\bibitem[{{Fouquet} {et~al.}(2012){Fouquet}, {hammer}, {Yang}, {Puech}, \&
  {Flores}}]{Fouquet2012}
{Fouquet}, S., {hammer}, H., {Yang}, Y., {Puech}, M., \& {Flores}, H. 2012,
  MNRAS

\bibitem[{{Hammer} {et~al.}(2009){Hammer}, {Flores}, {Puech}, {Yang},
  {Athanassoula}, {Rodrigues}, \& {Delgado}}]{Hammer2009}
{Hammer}, F., {Flores}, H., {Puech}, M., {et~al.} 2009, \aap, 507, 1313

\bibitem[{{Hammer} {et~al.}(2010){Hammer}, {Yang}, {Wang}, {Puech}, {Flores},
  \& {Fouquet}}]{Hammer2010}
{Hammer}, F., {Yang}, Y.~B., {Wang}, J.~L., {et~al.} 2010, \apj, 725, 542

\bibitem[{{Kroupa}(1997)}]{Kroupa1997}
{Kroupa}, P. 1997, \na, 2, 139

\bibitem[{{Kroupa} {et~al.}(2005){Kroupa}, {Theis}, \& {Boily}}]{Kroupa2005}
{Kroupa}, P., {Theis}, C., \& {Boily}, C.~M. 2005, \aap, 431, 517

\bibitem[{{Lynden-Bell}(1976)}]{Lynden-Bell1976}
{Lynden-Bell}, D. 1976, \mnras, 174, 695

\bibitem[{{Metz} {et~al.}(2007){Metz}, {Kroupa}, \& {Jerjen}}]{Metz2007}
{Metz}, M., {Kroupa}, P., \& {Jerjen}, H. 2007, \mnras, 374, 1125

\bibitem[{{Metz} {et~al.}(2009){Metz}, {Kroupa}, \& {Jerjen}}]{Metz2009}
{Metz}, M., {Kroupa}, P., \& {Jerjen}, H. 2009, \mnras, 394, 2223

\bibitem[{{Pawlowski} {et~al.}(2011){Pawlowski}, {Kroupa}, \& {de
  Boer}}]{Pawlowski2011}
{Pawlowski}, M.~S., {Kroupa}, P., \& {de Boer}, K.~S. 2011, \aap, 532, A118

\bibitem[{{Pawlowski} {et~al.}(2012){Pawlowski}, {Pflamm-Altenburg}, \&
  {Kroupa}}]{Pawlowski2012}
{Pawlowski}, M.~S., {Pflamm-Altenburg}, J., \& {Kroupa}, P. 2012, \mnras, 423,
  1109

\bibitem[{{Robotham} {et~al.}(2012){Robotham}, {Baldry}, {Bland-Hawthorn},
  {Driver}, {Loveday}, {Norberg}, {Bauer}, {Bekki}, {Brough}, {Brown},
  {Graham}, {Hopkins}, {Phillipps}, {Power}, {Sansom}, \&
  {Staveley-Smith}}]{Robotham2012}
{Robotham}, A.~S.~G., {Baldry}, I.~K., {Bland-Hawthorn}, J., {et~al.} 2012,
  \mnras, 424, 1448

\bibitem[{{van der Marel} {et~al.}(2012){van der Marel}, {Fardal}, {Besla},
  {Beaton}, {Sohn}, {Anderson}, {Brown}, \& {Guhathakurta}}]{VdM2012}
{van der Marel}, R.~P., {Fardal}, M., {Besla}, G., {et~al.} 2012, \apj, 753, 8

\bibitem[{{Walker} {et~al.}(2009){Walker}, {Mateo}, {Olszewski},
  {Pe{\~n}arrubia}, {Wyn Evans}, \& {Gilmore}}]{Walker2009}
{Walker}, M.~G., {Mateo}, M., {Olszewski}, E.~W., {et~al.} 2009, \apj, 704,
  1274

\bibitem[{{Wang} {et~al.}(2012){Wang}, {Frenk}, \& {Cooper}}]{Wang2012}
{Wang}, J., {Frenk}, C.~S., \& {Cooper}, A.~P. 2012, ArXiv e-prints

\bibitem[{{Wetzstein} {et~al.}(2007){Wetzstein}, {Naab}, \&
  {Burkert}}]{Wetzstein2007}
{Wetzstein}, M., {Naab}, T., \& {Burkert}, A. 2007, \mnras, 375, 805

\bibitem[{{Yang} \& {Hammer}(2010)}]{Yang2010}
{Yang}, Y. \& {Hammer}, F. 2010, \apjl, 725, L24

\end{thebibliography}

\end{document}